\documentclass[11pt]{article}
\usepackage{amssymb}
\usepackage{amsmath}
\usepackage{graphicx}
\usepackage{bm}
\newcommand{\mathsym}[1]{{}}

\usepackage{graphicx}
\usepackage{rotating}
\usepackage{color}
\usepackage{cite}

\usepackage{graphicx}  % needed for figures
\usepackage{dcolumn}   % needed for some tables
\usepackage{bm}        % for math
\usepackage{amssymb}   % for math
\usepackage{amsbsy}
\usepackage{amsmath}
\usepackage{amsfonts}
\usepackage{color}
\usepackage[dvipsnames]{xcolor}
\usepackage{amsthm}
\usepackage{graphics}
\usepackage{graphicx}
\usepackage{xcolor}
\usepackage{caption}
\usepackage{subcaption}
\usepackage{amssymb}

\theoremstyle{definition}
\theoremstyle{definition}

\theoremstyle{definition}	 %AÃ±adimos esto para que tenga el estilo de una definiciÃ³n y no la de teorema o proposiciÃ³n. Por defecto el estilo es "plain", que es el de teoremas.
\theoremstyle{definition}
\theoremstyle{definition} %AÃ±adimos esto para que tenga el estilo de una definiciÃ³n y no la de teorema o proposiciÃ³n. Por defecto el estilo es "plain", que es el de teoremas.

\theoremstyle{definition} %El estilo remark podrÃ­a ponerse aquÃ­ tambiÃ©n

\theoremstyle{definition} %El estilo remark podrÃ­a ponerse aquÃ­ tambiÃ©n

\makeatletter
\renewcommand{\p@subsection}{}
\renewcommand{\p@subsubsection}{}
\makeatother
\usepackage{xcolor}
\usepackage{graphicx}
\definecolor{title}{RGB}{0,0,0}
\definecolor{other}{RGB}{0,0,0}
\definecolor{name}{RGB}{0,0,0}
\definecolor{phd}{RGB}{0,0,0}
\usepackage[hidelinks]{hyperref}
\hypersetup{
  colorlinks   = true, 	%Colours links instead of ugly boxes
  urlcolor     = red, 	%Colour for external hyperlinks
  linkcolor    = blue, 	%Colour of internal links
  citecolor   = magenta 	%Colour of citations
}

\definecolor{applegreen}{rgb}{0.0, 0.71, 0.0}

\begin{document}

\vspace{20pt}

\begin{center}

{\Large\bf Two types of $q$-Gaussian distributions used to study the diffusion in a finite region}
\vspace{10pt}

{\large  Won Sang Chung${}^1$\footnote{mimip44@naver.com, ORCID: \href{http://orcid.org/0000-0002-1358-6384}{0000-0002-1358-6384}}, 
L. M. Nieto${}^2$\footnote{luismiguel.nieto.calzada@uva.es, 
ORCID: \href{http://orcid.org/0000-0002-2849-2647}{0000-0002-2849-2647}}, 
Soroush Zare${}^2$\footnote{szare@uva.es, ORCID: \href{http://orcid.org/0000-0003-0748-3386}{0000-0003-0748-3386}},
and Hassan Hassanabadi${}^{2}$\footnote{hha1349@gmail.com, ORCID: \href{http://orcid.org/0000-0001-7487-6898}{0000-0001-7487-6898}}}
\vspace{10pt}

{\sl ${}^{1}$Department of Physics and Research Institute of Natural Science, \\
 College of Natural Science,
 Gyeongsang National University, Jinju 660-701, Korea}
 \\ 
 \medskip
{\sl ${}^{2}$Departamento de F\'{\i}sica Te\'{o}rica, At\'{o}mica y \'{O}ptica, and Laboratory for Disruptive Interdisciplinary Science (LaDIS), Universidad de Valladolid, 47011 Valladolid, Spain}

\vspace{12pt}

\today

\end{center}

\begin{abstract}
In this work, we explore both the ordinary $q$-Gaussian distribution and a new one defined here, determining both their mean and variance, and we use them to construct solutions of the $q$-deformed diffusion differential equation. This approach allows us to realize that the standard deviation of the distribution must be a function of time. In one case, we derive a linear Fokker-Planck equation within a finite region, revealing a new form of both the position- and time-dependent diffusion coefficient and the corresponding continuity equation. It is noteworthy that, in both cases, the conventional result is obtained when $q$ tends to zero. Furthermore, we derive the deformed diffusion-decay equation in a finite region, also determining the position- and time-dependent decay coefficient.
A discrete version of this diffusion-decay equation is addressed, in which the discrete times have a uniform interval, while for the discrete positions the interval is not uniform.
\end{abstract}

\vspace{10pt}

\vspace{12pt}

\noindent
AUTHOR CONTRIBUTIONS: 
The authors state that the research was carried out in conjunction with one another and with equal accountability. The final text was reviewed and approved by all of the authors.
\vspace{12pt}

\noindent
CONFLICT OF INTEREST STATEMENT: 
The writers state unequivocally that they do not have any conflicts of interest.
\vspace{12pt}

\newpage

\section{Introduction}

The $q$-Gaussian probability distribution represents a generalized form of the Gaussian distribution that has been extensively studied \cite{TsallisJSP1988, TsallisPA2004,TsallisPTPS2006}, since it is of great interest because it is derived from an entropy function in the framework of non-extensive statistical mechanics.
The $q$-Gaussian distribution function finds numerous applications in various scientific fields, as evidenced in \cite{Gell-Mann2004} and also in \cite{Frisch1995}, where the theory of multifractals, which arises in the study of turbulence observations.
But these are by no means the only ones. For example, in \cite{BurlagaApJL2020} they were used to explore the magnetic field of the very local interstellar medium observed by the twin  Voyager~1 and Voyager~2 spacecraft, in \cite{WitkovskyMetrol2023} their applications in measurement and metrology are explored, in \cite{SicuroAP2015} the robustness of the $q$-Gaussian family as attractors is analyzed using three deformations: the $\alpha$-, $\beta$-, and $\gamma$-triangles. In \cite{daSilvaEPL2023} it has been shown that the time series of the gravitational waves follows the $q$-Gaussian Tsallis' distribution as a probability density, and its dynamics evolve from the three corresponding Tsallis' indices, called $q$-triplets.
In \cite{SekaniaPA2018} it is shown that Compton profiles can be modeled by a $q$-Gaussian distribution and in \cite{DinhEPJB2018} that some types of financial systems can be modeled by $q$-Gaussian cumulative distribution functions.
In \cite{Matsuzoe2021} an investigation of the gauge freedom of entropies in $q$-Gaussian measurements is carried out.
Entropic extension and large deviations in the presence of strong correlations using warped distributions are investigated in \cite{TirnakliPD2022}.
In \cite{PajaresEPJA2023}, the authors derived an analytical function to describe the entire transverse momentum spectrum by a $q$-Gaussian distribution and, based on the distribution, described the string tension fluctuations.
In \cite{KotaAP2022}, after introducing some $q$-deformed distributions, statistical nuclear spectroscopy is investigated.
And finally, let us mention that in \cite{LiApJL2023}, the authors studied the self-organized criticality in precursors of long gamma-ray bursts in the third Swift/Burst Alert Telescope catalog, and they inspected the cumulative distribution functions of the size differences with a $q$-Gaussian function.

If we now focus on the field of non-extensive entropy, also called $q$-entropy \cite{TsallisJSP1988,CuradoJPA1991}, which was first proposed by Tsallis \cite{TsallisJSP1988}, we see that two deformed functions called $q$-exponential and $q$-logarithm are introduced there, which can be written in the following way
\begin{equation}\label{eq1-6}
e_q (x) = 
\left\{
\begin{array}{cl}
	0, & ~\mbox{if}~1 + (1-q) x <0\\
 \ 	( 1 + (1-q) x)^{\frac{1}{1-q}}, & ~ \mbox{if}~ 1 + (1-q) x\ge 0, \\ [1.2ex]
\ 	e^x, & ~\mbox{if}~1-q=0, \\
 \ 	( 1 + (1-q) x)^{\frac{1}{1-q}}, & ~ \mbox{if}~ 1-q>0~ and ~1+(1-q)x<0, \\ [1.2ex]
\end{array}
\right.
\quad 
\ln_q x = \frac{x^{{\color{black}1-q}} -1}{{\color{black}1-q}}.
\end{equation}

Some applications of $q$-entropy to statistical physics and other related fields have been achieved in
\cite{Naudts2011,NobrePRL2022,DinizBJMBR2010,FerriPA2010,EsquivelApJ2010,SotolongoGrauPRL2010,AfsarPRE2010,ConroyPRD2008,PlastinoPLA1993,RigoPLA2000,AssisJMP2005,PlastinoAP1997,ManzanoNJP2010}. 
Furthermore, some authors \cite{NivanenRMP2003,BorgesPA2004} also carried out mathematical studies on $q$-exponential, where $q$-addition, $q$-subtraction, $q$-product and $q$-division were introduced, which were used to construct a $q$-derivative, which is different from another $q$-derivative \cite{JacksonMM1909} that appears in the $q$-boson theory \cite{ArikJMP1976,MacfarlaneJPAMG1989,BiedenharnJPAMG1989}. Using the $q$-addition, $q$-deformed quantum mechanics \cite{ChungFdP2019,ChungIJMP2019} was formulated so that it could possess a $q$-translational symmetry, described by the $q$-addition.
In \cite{NaudtsJIPAM2004,NaudtsEpL2005,NaudtsOSID2005,NaudtsPA2006,NaudtsE2008,NaudtsCEJP2009} some ideas about the family of exponential functions have been generalized, which have been expanded from the mathematical literature in \cite{NaudtsJIPAM2004,NaudtsE2008,Grunwald2004,Eguchi2006}. This extension has been investigated in \cite{AbeJPA2003,HanelPA2007} from the perspective of the principle of maximum entropy, and its subsequent application in game theory can be seen in Refs. \cite{Topsoe2007,Topsoe2009}.

In this context, it is absolutely natural to analyze possible generalizations of the standard Gaussian distribution, which as we all know plays a crucial role in both mathematics and physics, specifically to describe some notable physical situation such as diffusion processes, wave packets, etc. 
In \cite{UmarovMJM2008} a $q$-generalization of the standard Gaussian distribution was already introduced using the $q$-exponential given in \eqref{eq1-6} as $ P_q (x,a) \propto e_q ( - a x^2),~ a>0$,
and has been applied to various scientific fields, such as statistical mechanics, geology, astronomy, economics and machine learning \cite{BurlagaApJL2020,WitkovskyMetrol2023,SicuroAP2015,daSilvaEPL2023,SekaniaPA2018,DinhEPJB2018,Matsuzoe2021,TirnakliPD2022,PajaresEPJA2023,KotaAP2022,LiApJL2023}. For some values of $q$, the $q$-Gaussian distribution is the probability distribution function (PDF) of a bounded random variable, which makes in biology and other domains \cite{dOnofrio2013} this new $q$-Gaussian distribution may be more suitable than the original Gaussian distribution to model the effect of external stochasticity.

This work is organized as follows. In Section~\ref{2types} we will use this $q$-Gaussian distribution, which we will henceforth call the Type-1 $q$-Gaussian distribution, and we will also propose a second different form for the $q$-Gaussian distribution, which we will call Type~2 $q$-Gaussian distribution. Both will be used in Section~\ref{Diffusion}, which is the essential part of this work, to study the diffusion process in a finite region. The work will end with the conclusions of Section~\ref{conclusions}.

\section{Two types of $q$-Gaussian distributions}\label{2types}

\begin{figure}[htb]
	\centering % \begin{center}/\end{center} takes some additional vertical space
	\includegraphics[width=.44\textwidth]{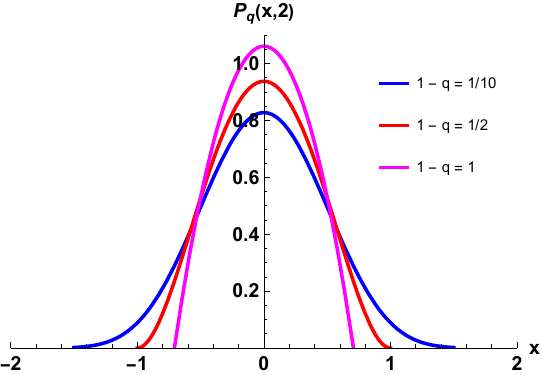}
	\hfill
	\includegraphics[width=.44\textwidth]{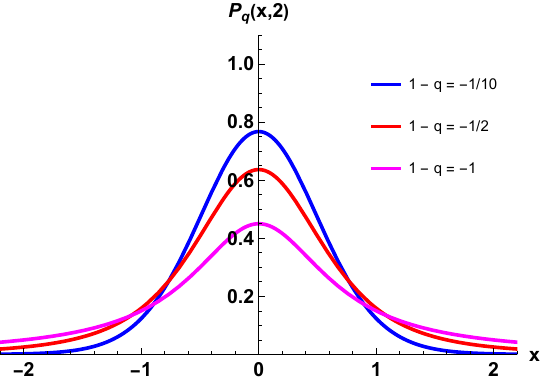}
	% "\includegraphics" is very powerful; the graphicx package is already loaded
	\caption{\label{fig:1} On the left, a graph illustrating $P_q(x,2)$ as a function of $x$ for three positive values of ${\color{black}1-q}$, which logically gives rise to distributions with compact support $|x|\leq 1/\sqrt{2({\color{black}1-q})}$. On the right, a similar graph showing $P_q(x,2)$ as a function of $x\in\mathbb R$ for three negative values of ${\color{black}1-q}$.}
\end{figure}

As already mentioned, starting from the definition of $q$-exponential given in the equation \eqref{eq1-6}, we will now consider two types of $q$-Gaussian distributions.
The first of them, which we will call $q$-Gaussian Type 1 distribution, has already been previously introduced:
\begin{equation}\label{eq2-1}
P_q (x,a) =  \sqrt{a}\, A_q \ e_q ( - a x^2) = \sqrt{a}\, A_q
\left\{
\begin{array}{cl}
\left( 1 - a\, ({\color{black}1-q})\,  x^2 \right)^{\frac{1}{{\color{black}1-q}}}, &\  \mbox{if}~~ 1\ge  a\, ({\color{black}1-q})\,  x^2, \\ [1.5ex]
\ \ \ \ \	0, & \  \mbox{if}~~1 <a\, ({\color{black}1-q})\,  x^2,
\end{array}
\right.
\quad 
 a>0.
\end{equation}
From the previous expression it is easily deduced that if ${\color{black}1-q}>0$, then the distribution $P_q (x,a)$ has compact support, which is the closed interval $|x|\leq 1/\sqrt{a({\color{black}1-q})}$, while if ${\color{black}1-q}\leq 0$, then the distribution $P_q (x,a)$ makes sense for all $x\in\mathbb R$, and the support is not compact.
The constant $A_q$ in \eqref{eq2-1} is obtained by normalizing the function, that is, by imposing $\int_{\mathbb R}P_q (x,a)\, dx=1$, and it is given by
\begin{equation}\label{eq2-6}
A_{q}=
\left\{
\begin{array}{cl}
\dfrac{\sqrt{{\color{black}1-q}}\ \Gamma \left(\frac{2+3({\color{black}1-q})}{2({\color{black}1-q})}\right)}{\sqrt{\pi}\ \Gamma \left(\frac{1+({\color{black}1-q})}{({\color{black}1-q})}\right)},
& {\text{ for }}~ {\color{black}1-q}>0 , \\  [3.5ex]
\dfrac1{\sqrt{\pi}},   &  {\text{ for }}~{\color{black}1-q}=0,   \\ [2ex]
\dfrac{\sqrt{|{\color{black}1-q}|}\ \Gamma \left(\frac1{|{\color{black}1-q}|}\right)}{\sqrt{\pi}\ \Gamma \left(\frac{2-|{\color{black}1-q}|}{2|{\color{black}1-q}|}\right)},
& {\text{ for }}~-2<{\color{black}1-q}<0,
\end{array}
\right.
\end{equation}
Therefore, for the distribution $P_q (x,a)$  to be normalizable, the parameter $q$ must be constrained by the condition ${\color{black}1-q}>-2$.
It can  also be verified that in the limit ${\color{black}1-q}\to 0$ we have
\begin{equation} 
\lim_{q\to 0} P_q (x,a) = \sqrt{\frac{a}\pi }\ e^{- a x^2}. 
\end{equation}
As already said, this distribution has been considered previously in the literature \cite{UmarovMJM2008} and Fig.~\ref{fig:1} represents the $q$-Gaussian Type-1 distribution for different values of the parameter $q$, both positive and negative. From the equation \eqref{eq2-1} it follows that the maximum value of this $q$-Gaussian distribution is obtained at $x=0$ and its value is given by $P_q (0,a) = \sqrt{a }\, A_q$.

It can be seen from the drawings that for both positive and negative values of ${\color{black}1-q}$, as ${\color{black}1-q}$ increases, the peak of the $q$-Gaussian distribution also increases, but the corresponding spread decreases. 
%On the contrary, this trend is reversed when we consider values $q<0$.
In fact, the mean and variance can be found and are given by
\begin{equation}\label{eq2-7}
E_1(X)=0
\quad
\text{and}
\quad
\sigma_1^2 =\langle x^2 \rangle
= \begin{cases} \dfrac{1}{ a(2 +3({\color{black}1-q})) }, \quad{\text{ for }}~{\color{black}1-q}> -\frac{2}{3},  \\
	\ \ \ \infty,\qquad  {\text{ for }}~ {\color{black}1-q} \le -\frac{2}{3}.
\end{cases}
\end{equation}
Therefore, the variance of $P_q(x,a)$ is  finite only if the condition ${\color{black}1-q}>-2/3$ is satisfied.

But this is not the only possibility of defining a deformed Gaussian.
Another interesting and useful possibility is what we will call $q$-Gaussian Type 2 distribution, whose probability distribution function for $a>0$ is 
\begin{equation}\label{eq2-9}
G_q (x,a) =B_q  \, [e_q ( - x^2)]^a= B_q  \, [e_{-q}(  x^2)]^{-a} = B_q 
\left\{
\begin{array}{cl}
\left( 1 - ({\color{black}1-q})\,  x^2 \right)^{\frac{a}{{\color{black}1-q}}}, &
\mbox{if}~~ 1\ge ({\color{black}1-q})\,  x^2, \\ [1.5ex]
\ 	0, & \mbox{if}~~1 < ({\color{black}1-q})\,  x^2,
\end{array}
\right. 
\end{equation}
From here we can see that if ${\color{black}1-q}>0$, then the distribution $G_q (x,a)$ has a compact support, which is the closed interval $|x|\leq 1/\sqrt{{\color{black}1-q}}$, while if ${\color{black}1-q}\leq 0$ the distribution $G_q (x,a)$ exists for all $x\in\mathbb R$.
The constant $B_q$ is obtained by normalizing the function $G_q (x,a)$, $\int_{\mathbb R}G_q (x,a)\, dx=1$, and it is given by
\begin{equation}\label{eq2-10}
B_q =
\left\{
\begin{array}{cl}
\dfrac{\sqrt{{\color{black}1-q}}\ \Gamma \left(\frac{2a+3({\color{black}1-q})}{2({\color{black}1-q})}\right)}{\sqrt{\pi}\ \Gamma \left(\frac{a+({\color{black}1-q})}{{\color{black}1-q}}\right)},
& {\text{ for }}~ {\color{black}1-q}>0 , \\  [3.5ex]
\sqrt{\dfrac{a}{\pi}},   &  {\text{ for }}~{\color{black}1-q}=0,   \\ [2ex]
\dfrac{\sqrt{|{\color{black}1-q}|}\ \Gamma \left(\frac{a}{|{\color{black}1-q}|}\right)}{\sqrt{\pi}\ \Gamma \left(\frac{2a-|{\color{black}1-q}|}{2|{\color{black}1-q}|}\right)},
& {\text{ for }}~-2a<{\color{black}1-q}<0.
\end{array}
\right.
\end{equation}
From \eqref{eq2-9} it follows that the maximum value of this $q$-Gaussian distribution is obtained at $x=0$ and its value is given by $G_q (0,a) = B_q$.

It can  also be verified that in the limit ${\color{black}1-q}\to 0$ we have
\begin{equation} 
\lim_{q\to 0} G_q (x,a) = \sqrt{\frac{a}\pi }\ e^{- a x^2},
\end{equation}
for which Stirling's formula is used.
The mean and variance of the $q$-Gaussian Type 2 distribution are given by
\begin{equation}\label{eq2-11}
E_2(X)=0
\quad
\text{and}
\quad
\sigma_2^2 =\langle x^2 \rangle
= \begin{cases} \dfrac{1}{ 2a +3({\color{black}1-q}) }, ~~ {\color{black}1-q}>-\frac{2a}3  ,\\
	~~\infty, ~~~~{\color{black}1-q}\leq -\frac{2a}3.
\end{cases}
\end{equation}

\begin{figure}[h]
	\centering % \begin{center}/\end{center} takes some additional vertical space
	\includegraphics[width=.44\textwidth]{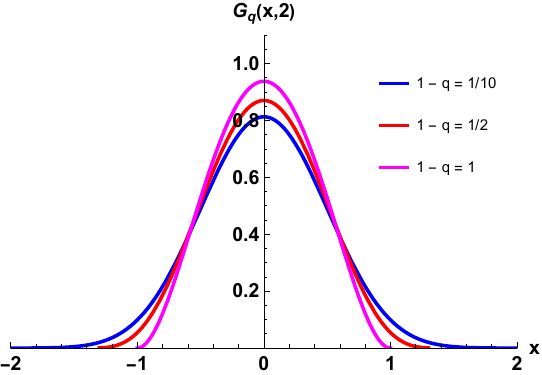}
	\hfill
	\includegraphics[width=.44\textwidth]{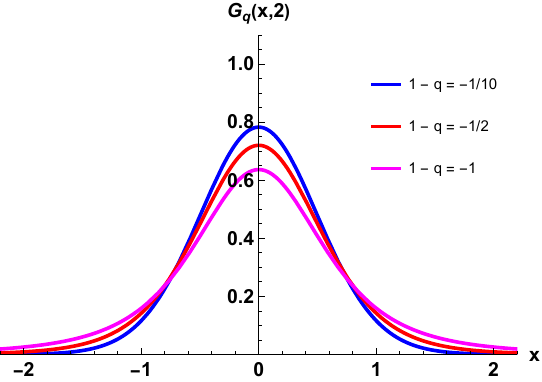}
	% "\includegraphics" is very powerful; the graphicx package is already loaded
	\caption{\label{fig:2} On the left, plot of $G_q(x,2)$ as a function of $x$ for the three positive values of ${\color{black}1-q}$ indicated.
	On the right, a similar graph showing $G_q(x,2)$ as a function of $x$ for three negative values of ${\color{black}1-q}$.
	The presence of a different compact support for each ${\color{black}1-q}>0$ is clearly seen on the left plot, while on the right the support of all the functions is the whole real line. 
	}
\end{figure}

To illustrate the behavior of the functions $G_q(x,a)$, Fig.~\ref{fig:2} shows some of them for $a=2$ and various values of the parameter ${\color{black}1-q}$.
Figure \ref{fig:2} shows how the peak of this $q$-Gaussian distribution for ${\color{black}1-q}$ increases with growth of ${\color{black}1-q}$, while the associated dispersion falls.
Finally, it is worth noting that the two $q$-Gaussian distributions are related as follows:
\begin{equation}
G_{aq}(x,a)=P_{q}(x,a).
\end{equation}
In fact, there are not just two possibilities, but an infinite family of such distributions: $E_{(q,\gamma)}(x,a)= [1-a^{\gamma} (1-q) x^2]^{a^{1-\gamma}/(1-q)}$. 
The two cases we have analyzed above, corresponding to $\gamma=1$ (Type 1) and $\gamma=0$ (Type 2), are particularly interesting: Type 1 can be derived from the maximum $q$-entropy principle, and Type 2 can be derived from the  $q$-deformed spacial derivative ${\cal D}_x^q$, which has the $q$-deformed translation symmetry $x \rightarrow x \oplus \delta x$. 
The $q$-deformed translation symmetry is related to the $q$-sum of the $q$-deformed lattice. Type 1 is known to be related to Type 2 through property A.28 in Ref.~\cite{new}, and as $q$ approaches one, both Type 1 and Type 2 distributions become the Gaussian distribution.

\section{Diffusion process  in finite region}\label{Diffusion}

The standard diffusion equation in one spatial dimension is given by
\begin{equation}\label{eq3-0-1}
\partial_t \varrho(x, t) = D\, \partial_x^2\varrho ( x,t),
\end{equation}
where $\varrho(x,t)$ is the density of the diffusing material at location $x$ and time $t$, and $D$ is the diffusion coefficient, which depends on both the diffusing material and the substrate in which it diffuses and which we assume to be constant. A typical and simple initial condition used to solve \eqref{eq3-0-1} is
\begin{equation}\label{eq3-0-2}
\varrho(x, 0) =  \delta (x),
\end{equation}
where it has been established that the total mass of the substance released per unit cross-sectional area is unity. If we consider an infinite domain $(-\infty, \infty)$, the solution of \eqref{eq3-0-1} turns out to be
\begin{equation}\label{eq3-0-3}
\varrho(x, t) = \frac{1}{\sqrt{ 4 \pi \,D\, t}}\ e^{ -\frac{x^2}{4\,D\, t}} = \frac{1}{\sqrt{ 2 \pi}\, \sigma}\ e^{ -\frac{x^2}{2\,\sigma^2}}    .
\end{equation}
Note that the solution found in \eqref{eq3-0-3} is a Gaussian distribution whose mean is zero and whose variance is $\sigma^2 =\langle x^2 \rangle = 2 \,D\, t $. 
In the result \eqref{eq3-0-3} the conservation of the total amount of the substance that diffuses has already been taken into account, that is,\begin{equation}\label{eq3-0-4}
\int_{-\infty}^{\infty} \varrho(x, t)\,  dx = 1.
\end{equation}
The density $\varrho(x, t)$ has the following limits
\begin{equation}\label{eq3-0-5}
\lim_{x\rightarrow \pm \infty} \varrho(x, t) =0 \qquad \text{and}\qquad 
\lim_{t\rightarrow  \infty} \varrho(x, t) =0.
\end{equation}

In the following subsections we will discuss the solution of the diffusion equation \eqref{eq3-0-1} in a finite region $-L \le x \le L$, based on the $q$-Gaussian distributions with finite support that were previously introduced.

\subsection{Diffusion in a finite region based on Type 1 $q$-Gaussian with ${\color{black}1-q}>0$}

If we want to find a solution to the diffusion equation in a finite interval, an interesting idea is to use $q$-Gaussians with ${\color{black}1-q}>0$ which, as we have seen in the previous section, have a compact support and, therefore, automatically satisfy homogeneous boundary conditions, both at the ends and on the outside of the aforementioned compact where the solution is not zero.
Specifically, if we look for the concentration as a Type 1 $q$-Gaussian distribution with ${\color{black}1-q}>0$, we can use the fact that, as already seen, for this $q$-Gaussian the support is given by $ L = 1/\sqrt{a({\color{black}1-q})q}$, therefore we can take this value as a starting point to solve the problem.
Let us now assume that $\varrho_q(x,t)$ is the $q$-deformed density of the diffusing material at location $x$ and time $t$, the variance of which depends on the variables $x$, $t$ and the parameter ${\color{black}1-q}$. Within the support we can assume the following ansatz based on \eqref{eq3-0-3} and \eqref{eq2-1}:
\begin{eqnarray}  
\varrho_q(x, t) \!\!&\!=\!&\!\! \frac{K_q/\sqrt{ 2 + 3 ({\color{black}1-q})}}{\sigma_q(x,t) } e_q \left( - \frac{ x^2}{ (2+3({\color{black}1-q})) \sigma^2_q(x,t)}\right)
\nonumber \\ [2ex]
 \!\!&\!=\!&\!\!  \frac{K_q/\sqrt{ 2 + 3 ({\color{black}1-q})}}{\sigma_q(x,t) } 
\left[ 1- \frac{({\color{black}1-q}) x^2}{ (2+3({\color{black}1-q})) \sigma^2_q(x,t)}\right]^{\frac{1}{({\color{black}1-q})}} ,
\label{eq3-1-28--Second}
\end{eqnarray}
where $\sigma^2_q(x,t)$ is a standard deviation $\langle x^2\rangle$ for this $q$-deformed density \eqref{eq3-1-28--Second}, that obviously will depend on ${\color{black}1-q}$, $x$ and $t$, and the range of $x$ is
\begin{equation}\label{eq3-1-1}
- \sigma_q(x,t) \sqrt{ \frac{2+ 3({\color{black}1-q})}{{\color{black}1-q}}} \le x \le \sigma_q(x,t) \sqrt{ \frac{2+ 3({\color{black}1-q})}{{\color{black}1-q}}}.
\end{equation}
In a finite region $-L \le x \le L$ we can not demand $ \langle x^2\rangle= 2 Dt$ because we would obtain the undesirable result that $ \lim_{t\rightarrow \infty} \langle x^2\rangle= \infty $. Instead we can assume that
\begin{equation}\label{eq3-1-2}
\langle x^2\rangle =\sigma^2_q(x,t) = \frac{1}{{\color{black}1-q}} ( 1 - e^{ - 2 ({\color{black}1-q}) Dt}),
\end{equation}
which reduces to $\langle x^2\rangle= 2 Dt$ in the limit ${\color{black}1-q} \rightarrow 0$, as it should be. For Eq. \eqref{eq3-1-2} we have the limit
\begin{equation}\label{eq3-1-3}
\lim_{t \rightarrow \infty} \langle x^2\rangle = \frac{1}{{\color{black}1-q}},
\end{equation}
which is finite unless ${\color{black}1-q}=0$. Therefore, we can say that the diffusion characterized by the concentration \eqref{eq3-1-28--Second} has a moving boundary because $\sigma_q(x,t)$ is time-dependent in Eq. \eqref{eq3-1-2}. In the limit $t \rightarrow \infty$, the domain of $x$ is time idependent:
\begin{equation}\label{eq3-1-4}
-\frac{ \sqrt{ 2+ 3({\color{black}1-q})}}{{\color{black}1-q}} \le x \le \frac{ \sqrt{ 2+ 3({\color{black}1-q})}}{{\color{black}1-q}}.
\end{equation}
If we consider the diffusion in $|x|\le L$, we should demand
\begin{equation}\label{eq3-1-5}
\frac{ \sqrt{ 2+ 3({\color{black}1-q})}}{{\color{black}1-q}} \le L,
\end{equation}
which gives the lower bound for ${\color{black}1-q}$,
\begin{equation}%\label{eq3-1-6}
({\color{black}1-q}) \ge \frac{ 3 + \sqrt{ 9 + 8 L^2}}{ 2 L^2},
\end{equation}
Differentiating the concentration with respect to $t$ and $x$ leads to
\begin{equation}\label{eq3-1-6}
\frac{\partial \varrho_q (x, t)}{\partial t }  
= -\frac{D( 1 -({\color{black}1-q}) \sigma^2_q(x,t))}{ \sigma^2} \left[ 1 - \frac{2 x^2}{ (2 + 3 ({\color{black}1-q})) \sigma^2_q(x,t)} \left( \frac{1}{  1- \frac{({\color{black}1-q}) x^2}{ ( 2 + 3 ({\color{black}1-q}) ) \sigma^2_q(x,t)}}\right) \right]\!\! \varrho_q(x, t),
\end{equation}
and differentiating the equation \eqref{eq3-1-28--Second} with respect to $t$ and $x$ we arrive at
\begin{equation}\label{eq3-1-7}
\frac{\partial}{\partial x } \varrho_q (x, t)
= -\frac{2 x}{ (2 + 3 ({\color{black}1-q})) \sigma^2_q(x,t)} \left( \frac{1}{  1- \frac{({\color{black}1-q}) x^2}{ ( 2 + 3 ({\color{black}1-q}) ) \sigma^2_q(x,t)}}\right)  \varrho_q(x, t),
\end{equation}
where we have used the following equation obtained from \eqref{eq3-1-2}
\begin{equation}\label{eq3-1-8}
\sigma_q(x,t) \frac{d\sigma_q(x,t)}{dt} = D ( 1 - ({\color{black}1-q}) \sigma^2).
\end{equation}
Therefore, we have a linear Fokker-Planck equation in a finite region:
\begin{equation}\label{eq3-1-9}
\frac{\partial}{\partial t } \varrho_q(x, t) = \frac{\partial}{\partial x }\left( D_q (x, t) \frac{\partial}{\partial x } \varrho_q(x, t) \right),
\end{equation}
where the position- and time-dependent diffusion coefficient is
\begin{equation}\label{eq3-1-10}
D_q (x, t) = D \left( 1 + \frac{3}{2} ({\color{black}1-q})\right) ( 1 - ({\color{black}1-q}) \sigma^2_q(x,t)) \left(1- \frac{({\color{black}1-q}) x^2}{ ( 2 + 3 ({\color{black}1-q}) ) \sigma^2_q(x,t)}\right),
\end{equation}
which reduces to $D$ in the limit ${\color{black}1-q}\rightarrow 0$.
In fact we have shown that the diffusion differential equation changes to the linear Fokker-Planck equation by changing the density to the $q$-deformed density and the constant diffusion coefficient to a position- and time-dependent diffusion coefficient.

\begin{figure}[htb]
	\centering % \begin{center}/\end{center} takes some additional vertical space
	\centering % \begin{center}/\end{center} takes some additional vertical space
\includegraphics[width=.44\textwidth]{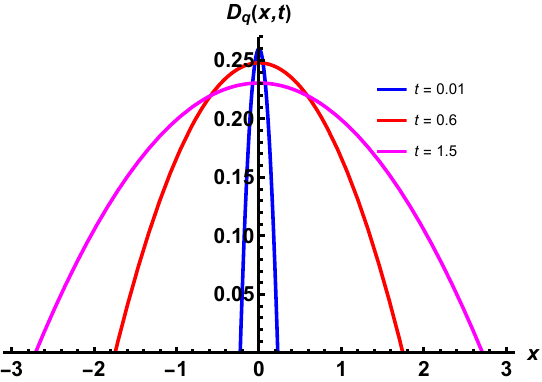}
\hfill
\includegraphics[width=.44\textwidth]{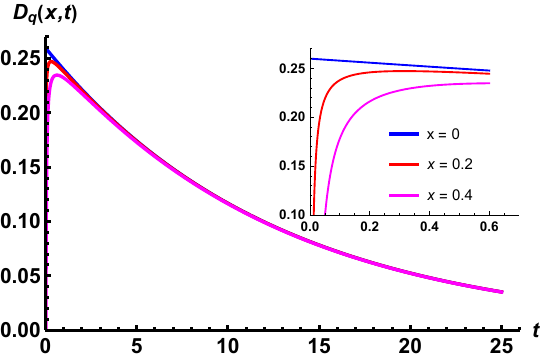}
	\caption{\label{fig:3}  
		On the left, plots of $D_q(x,t)$ for three different values of $t$.
		On the right, plots of $D_q(x,t)$ for three different values of $x$. In all cases we have chosen $D={\color{black}1-q}=0.2$.
	}
\end{figure}

In the left panel of Fig.~\ref{fig:3}, we illustrate the graph of $D_q(x,t)$ as a function of $x$ for three different values of $t$, and in the right panel we illustrate shows the graph of $D_q(x,t)$ in terms of $t$ for three different values of $x$.
In all scenarios, as $q$ goes to zero,  the position- and time-dependent coefficient converges to a constant value, independent of time and position. During the initial stages of the diffusion scattering phenomenon, the diffusion coefficients first increase and then decrease. Furthermore, regardless of the values of ${\color{black}1-q}$ and $x$, as $t$ tends toward infinity, the position- and time-dependent coefficient stabilizes at a constant value, which is not affected by the parameters.

To finish, remark that Eq. \eqref{eq3-1-9} is the continuity equation
\begin{equation}\label{eq3-1-11}
\frac{\partial }{\partial t}  \varrho_q (x, t) + \frac{\partial }{\partial x} J_q (x, t) =0,
\end{equation}
where the flux $J_q (x, t)$ obeys Fick's law \cite{Fick1855}
\begin{equation}\label{eq3-1-12}
J_q (x, t) = - D_q (x, t)  \frac{\partial }{\partial x} \varrho_q (x, t).
\end{equation}

\subsubsection{Case in which a substance is released at $x=x_0$ }

If a substance is released at $x=x_0$ ( $-L < x_0 < L$), then the $q$-deformed density of the diffusing material in Eq. \eqref{eq3-1-28--Second} becomes
\begin{eqnarray}
\varrho_q(x, t) \!\!&\!=\!&\!\!\frac{K_q/\sqrt{ 2 + 3 ({\color{black}1-q})}}{\sigma_q(x,t)}e_q \left( - \frac{ (x- x_0)^2}{ ( 2 + 3 ({\color{black}1-q}) ) \sigma^2_q(x,t)}\right)  \nonumber \\ [2ex]
\!\!&\!=\!&\!\! 
 \frac{K_q/\sqrt{ 2 + 3 ({\color{black}1-q})}}{\sigma_q(x,t)} \left[ 1- \frac{({\color{black}1-q}) (x-x_0)^2}{ ( 2 + 3 ({\color{black}1-q}) ) \sigma^2_q(x,t)}\right]^{\frac{1}{{\color{black}1-q}}},
\label{eq3-1-13}
\end{eqnarray}
The domain of $x$ is then given by
\begin{equation}\label{eq3-1-14}
- L \le x_0 - \frac{ \sqrt{ 2+ 3({\color{black}1-q})}}{{\color{black}1-q}} \le x \le  x_0 + \frac{ \sqrt{ 2+ 3({\color{black}1-q})}}{{\color{black}1-q}} \le L,
\end{equation}
from which we obtain a lower bound for ${\color{black}1-q}$:
\begin{equation}\label{eq3-1-15}
{\color{black}1-q} \ge \frac{ 3 + \sqrt{ 9 + 8 (L-|x_0|)^2}}{ 2 (L-|x_0|)^2}.
\end{equation}

\subsection{Diffusion in a finite region based on Type 2 $q$-Gaussian with ${\color{black}1-q}>0$}
%For Type I $q$-Gaussian distribution with $q>0$ we have $x_{max} = \frac{1}{\sqrt{qa}}$, hence we can set $ L = \frac{1}{\sqrt{qa}}$.

\begin{figure}[htb]
	\centering % \begin{center}/\end{center} takes some additional vertical space
\includegraphics[width=.44\textwidth]{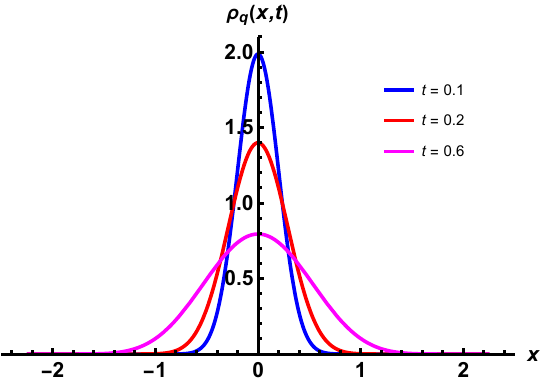}
\hfill
\includegraphics[width=.44\textwidth]{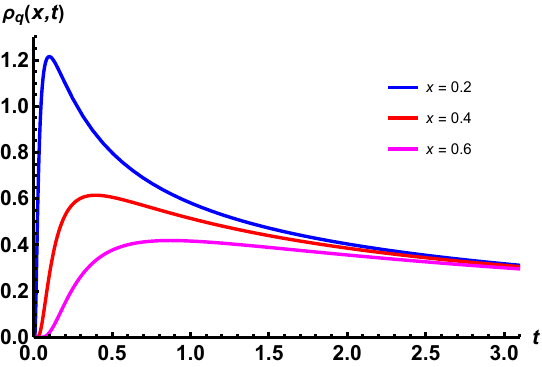}
	\caption{\label{fig:4}  
		On the left, plots of $\varrho_q(x,t)$ for three different values of $t$.
		On the right, plots of $\varrho_q(x,t)$ for three different values of $x$. In all cases we have chosen $D={\color{black}1-q}=0.2$.	Observe that in all cases $L\approx 2.2$. }
\end{figure}

Let us now consider the concentration as a $q$-Gaussian Type 2 distribution with ${\color{black}1-q}>0$.
Then, applying Eq.~\eqref{eq2-11}, we find the following function as a solution to the $q$-distorted diffusion differential equation: 
\begin{equation}\label{eq3-2-1}
\frac{1}{ 2 a(t) + 3 ({\color{black}1-q}) } = \sigma^2_q(x,t) =  \frac{1}{{\color{black}1-q}} ( 1 - e^{ - 2 ({\color{black}1-q}) Dt}),
\end{equation}
which gives
\begin{equation}\label{eq3-2-2}
a(t) = \frac{{\color{black}1-q}}{4} \left( \coth ( ({\color{black}1-q}) Dt) -5 \right).
\end{equation}
Thus, the concentration reads
\begin{equation}\label{eq3-2-3}
\varrho_q(x, t) = B_q (t) [ e_q (-x^2)]^{a (t)}.
\end{equation}
In this case we have
\begin{equation}\label{eq3-2-4}
L = \frac{1}{\sqrt{{\color{black}1-q}}}.
\end{equation}
In the left panel of Fig.~\ref{fig:4}, we represent $\varrho_q(x,t)$ as a function of $x$ for varying values of $t$, and here it can be seen that for increasing values of $t$, the dispersion of the distribution function widens.

In the right panel of Fig.~\ref{fig:4}, we depict the variation of $\varrho_q(x,t)$ with respect to $t$ for various values of $x$. Notably, the density dispersion exhibits an initial increase followed by a decrease, and as $t$ goes to infinity, $\varrho_q (x,t)$ converges to zero.

Differentiating the concentration with respect to $t$ and $x$, we have
\begin{equation}\label{eq3-2-5}
\frac{\partial}{\partial t } \varrho_q (x, t)
= \left( \frac{ \dot{B}_q(t)}{B_q(t)} + \frac{\dot{a}(t) \ln (1 - ({\color{black}1-q}) x^2)}{{\color{black}1-q}} \right)  \varrho_q(x, t),
\end{equation}
and
\begin{equation}\label{eq3-2-6}
\frac{\partial}{\partial x } \varrho_q (x, t)
= \left( -\frac{2 a(t) x}{ 1- ({\color{black}1-q}) x^2 } \right) \varrho_q(x, t).
\end{equation}
Hence, we have the deformed diffusion-decay equation in a finite region as
\begin{equation}\label{eq3-2-7}
\frac{\partial}{\partial t } \varrho_q(x, t) =  D ( {\cal D}_x^q)^2 \varrho_q (x, t) + \mu (x, t) \varrho_q (x, t),
\end{equation}
where the position- and time-dependent decay coefficient is
\begin{equation}\label{eq3-2-8}
\mu (x, t)  = \frac{ \dot{B}_q(t)}{B_q(t)} + \frac{\dot{a}(t) \ln (1 - ({\color{black}1-q}) x^2)}{{\color{black}1-q}}
+ 2 a D (1-({\color{black}1-q}) x^2) - 4 D a^2 x^2 ,
\end{equation}
and the operator ${\cal D}_x^q$, called the $q$-deformed derivative, is given by
\begin{equation}\label{eq3-2-9}
{\cal D}_x^q = ( 1 -({\color{black}1-q}) x^2)\frac{\partial}{\partial x}.
\end{equation}
From here we know that the position- and time-dependent decay  coefficient vanishes in the limit ${\color{black}1-q}\to 0$.

\subsubsection{Meaning of the deformed diffusion-decay equation in a discrete version}

Equation \eqref{eq3-2-7} is different from the ordinary diffusion-decay equation because the ordinary derivative with respect to $x$ is replaced with the deformed derivative ${\cal D}_x^q$. 
Now we will look for the physical meaning of the deformed diffusion-decay equation. 
The second term on right hand side of Eq. \eqref{eq3-2-7} represents the decay with position- and time-dependent decay  coefficient, while the first term differs from the ordinary diffusion term due to the presence of the deformed derivative.

Now let us consider the discrete version of Eq. \eqref{eq3-2-7} as
\begin{equation}\label{eq3-3-1}
\Delta_{\tau} \varrho_q ( x_n, t_m)
= D \Delta_{a}^2 \varrho_q ( x_n, t_m)  + \mu( x_n, t_m) \varrho_q (x_n , t_m),
\end{equation}
where time difference operator and position difference operator are given by
\begin{equation}\label{eq3-3-2}
\Delta_{\tau} F(x_n, t_m) = \frac{ F(x_n, t_{m+1}) -  F(x_n, t_{m})}{\tau},
\end{equation}
\begin{equation}\label{eq3-3-3}
\Delta_{a} F(x_n, t_m) = \frac{ F(x_{n+1}, t_{m}) -  F(x_n, t_{m})}{a}.
\end{equation}
If we consider the uniform time lattice obeying $t_{n+1}- t_n=\tau$ and take the limit $\tau\to 0$, then the time difference operator reduces to the ordinary time derivative. 
But this is not the case for the part corresponding to the spatial coordinate. The deformed derivative \eqref{eq3-2-9} implies that the position space must be discrete with a non-uniform interval. Here we assume that the discrete positions obey
\begin{equation}\label{eq3-3-4}
x_{n+1} \ominus x_n =a,
\end{equation}
where $\ominus$ is a deformed subtraction. 
The deformed derivative ${\cal D}_x^q$ given in \eqref{eq3-2-9} is not invariant under the ordinary translation $x \rightarrow x + a$ due to the presence of the factor $(1-({\color{black}1-q})x^2)$. 
Therefore, we will look for a $q$-deformed sum such that the $q$-deformed derivative \eqref{eq3-2-9} is invariant under the $q$-deformed sum. This can be achieved with the help of the pseudo-calculus \cite{PapNSJM1993} (or $f$-deformation \cite{ChungRMP2020}). Now let us consider $q$-deformed addition and $q$-deformed subtraction as
\begin{equation}\label{eq3-3-5}
a \oplus b = \frac{a +b }{ 1 + ({\color{black}1-q})  ab },
\qquad 
a \ominus b = \frac{ a -b} { 1 - ({\color{black}1-q})  ab }.
\end{equation}
Then, we have
\begin{equation}\label{eq3-3-7}
x_{n+1} = \frac{x_n +a  }{ 1 + ({\color{black}1-q})  ax_n  }.
\end{equation}
The action of the position $q$-difference operator on a function $F(x_n)$ is
\begin{equation}%\label{eq3-3-8}
\Delta_a F(x_n)
= \frac{1}{a} \left( F \left(  \frac{x_n +a  }{ 1 + ({\color{black}1-q})  ax_n  }\right) - F(x_n)\right).
\end{equation}
Now let us consider the case where $x_n$ corresponds to $x$ in the limit $a \rightarrow 0$. Then we set
\begin{equation}\label{eq3-3-8}
F \left(  \frac{x +a  }{ 1 +({\color{black}1-q})  ax  }\right)
= b_0 + b_1 a + {\cal O}(a^2),
\end{equation}
where
\begin{equation}\label{eq3-3-9}
b_0 = F(x),
\end{equation}
and
\begin{equation}\label{eq3-3-10}
b_1 = \left. \frac{\partial}{\partial a} F \left( \frac{x +a  }{ 1 + ({\color{black}1-q}) ax  }\right) \right|_{a=0} = ( 1 - ({\color{black}1-q}) x^2) \partial_x F(x).
\end{equation}
Therefore, the difference operator \eqref{eq3-3-3} reduces to the deformed derivative in the continuum limit. Thus, Eq. \eqref{eq3-2-7} is the continuum version of the  diffusion-decay equation where the discrete times have a uniform interval, while the discrete positions have a non-uniform interval defined in \eqref{eq3-3-4}.

\subsubsection{Case in which a substance is released at $x=x_0$}

\begin{figure}[htb]
	\centering % \begin{center}/\end{center} takes some additional vertical space
	\includegraphics[width=.48\textwidth]{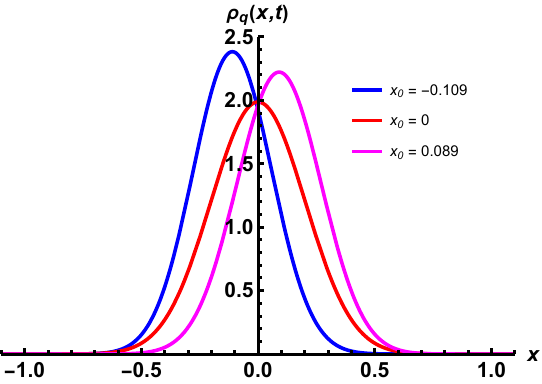}
	\caption{\label{fig:5}  Graph of the density $\varrho_q(x,t)$ given in \eqref{eq3-3-11} as a function of $x$ for three different values of $x_{0}$.  In all cases we have chosen $t=0.1$, $D={\color{black}1-q}=0.2$, so that $L\approx 2.2$.
	}
\end{figure}

Now let us consider the case where the substance to be diffused is released at $x=x_0$. If we impose $\varrho_q( \pm L)=0$, we can assume that the concentration takes the form
\begin{equation}\label{eq3-3-11}
\varrho_q (x, t) = b_q ( 1 + \sqrt{{\color{black}1-q}}\, x)^{\alpha} ( 1 - \sqrt{{\color{black}1-q}}\,  x)^{\beta}.
\end{equation}
The normalization is obtained from
\begin{equation}\label{eq3-3-12}
\int_{-\frac{1}{\sqrt{{\color{black}1-q}}}}^{\frac{1}{\sqrt{{\color{black}1-q}}}}  \varrho_q (x, t) \, dx =1,
\end{equation}
which gives
\begin{equation}\label{eq3-3-13}
b_q^{-1} = \frac{2^{\alpha +\beta +1} }{\sqrt{{\color{black}1-q}}} \left( \frac{ \Gamma ( \alpha +1) \Gamma (\beta +1)}{\Gamma ( \alpha + \beta +2)}\right).
\end{equation}
The average value is required to satisfy
\begin{equation}\label{eq3-3-14}
\langle x \rangle = \int_{-\frac{1}{\sqrt{{\color{black}1-q}}}}^{\frac{1}{\sqrt{{\color{black}1-q}}}}  x\, \varrho_q (x, t)\, dx = x_0.
\end{equation}
For the case ${\color{black}1-q}=0$ we have $\langle x^2 \rangle = x_0^2 + 2 Dt$, which gives
\begin{equation}\label{eq3-3-15}
\lim_{t \rightarrow \infty} \langle x^2 \rangle  = \infty.
\end{equation}
For ${\color{black}1-q}\ne 0$, we demand
\begin{equation}\label{eq3-3-16}
\lim_{t \rightarrow \infty} \langle x^2 \rangle  = \frac{1}{{\color{black}1-q}},
\end{equation}
which implies that the variance is given by
\begin{equation}\label{eq3-3-17}
\sigma^2_q(x,t) = \frac{1}{{\color{black}1-q}} ( 1 - e^{- 2 ({\color{black}1-q}) Dt}) -x_0^2.
\end{equation}
From the equations \eqref{eq3-3-14} and \eqref{eq3-3-17} the two unknowns $\alpha$ and $\beta$ can be found. Indeed, from \eqref{eq3-3-14} we have
\begin{equation}\label{eq3-3-18}
\frac{ \alpha - \beta}{ \sqrt{{\color{black}1-q}} ( \alpha + \beta +2)} = x_0,
\end{equation}
which gives
\begin{equation}\label{eq3-3-19}
\alpha = \left( \frac{ 1 + \sqrt{{\color{black}1-q}}\, x_0}{ 1 -\sqrt{{\color{black}1-q}}\, x_0} \right) \beta + \frac{ 2 \sqrt{{\color{black}1-q}}\, x_0}{ 1 - \sqrt{{\color{black}1-q}}\, x_0},
\end{equation}
and from Eq. \eqref{eq3-3-17}, we get
\begin{equation}\label{eq3-3-20}
\frac{1}{{\color{black}1-q}} \left( 1 - \frac{4 (\alpha+1)(\beta+1)}{( \alpha + \beta + 3)(\alpha + \beta +2)}  \right) = x_0^2 +\sigma^2_q(x,t).
\end{equation}
Solving \eqref{eq3-3-19}--\eqref{eq3-3-20} we get
\begin{eqnarray}
\alpha \!\!&\!=\!&\!\! \frac{ 1 }{ 4 ({\color{black}1-q}) \sigma^2_q(x,t)}
\Bigl[ 1 + \sqrt{{\color{black}1-q}}\,  x_0 - ({\color{black}1-q}) x_0^2 - 5 ({\color{black}1-q})\sigma^2_q(x,t)   \nonumber  \\
&& 
\!\! - ({\color{black}1-q}) \sqrt{{\color{black}1-q}}\,  x_0 ( \sigma^2_q(x,t) + x_0^2 ) + ( 1 + \sqrt{{\color{black}1-q}}\,  x_0) ( 1 - ({\color{black}1-q}) ( \sigma^2_q(x,t) + x_0^2 ))\Bigr],
\label{eq3-3-21}
\\
\beta  \!\!&\!=\!&\!\! \frac{ 1 }{ 4 ({\color{black}1-q}) \sigma^2_q(x,t)}
\Bigl[ 1 - \sqrt{{\color{black}1-q}}\,  x_0 - ({\color{black}1-q}) x_0^2 - 5 ({\color{black}1-q}) \sigma^2_q(x,t)  \nonumber  \\
&&  \!\! 
+ ({\color{black}1-q}) \sqrt{{\color{black}1-q}}\,  x_0 ( \sigma^2_q(x,t) + x_0^2 )+ ( 1 - \sqrt{{\color{black}1-q}}\,  x_0) ( 1 - ({\color{black}1-q}) (\sigma^2_q(x,t) + x_0^2 )) \Bigr].
\label{eq3-3-22}
\end{eqnarray}
It can be easily verified that for $x_0=0$ the equations \eqref{eq3-3-21} and \eqref{eq3-3-22} reduce to $ \alpha = \beta = a(t)/{\color{black}1-q}$, which corresponds to the equation \eqref{eq3-2-3}.
In Fig.~\ref{fig:5}, we plot the density $\varrho_q(x,t)$ given in \eqref{eq3-3-11} versus $x$ for different values of $x_0$.

\section{Concluding remarks}\label{conclusions}

In this article, we consider the $q$-Gaussian distributions of Type 1 and Type 2, obtained from two different definitions arising from the deformed $q$-exponential and the fact that $e_{q}(-ax^{2})\neq (e_{q}(-x^{2}))^{a}$.

We applied the Type 1 $q$-Gaussian distribution to find solutions to the diffusion equation and obtained the range of the standard deviation in this case.
Next we consider a special form for the $q$-deformed standard deviation, which reduces to the ordinary case when a long time is allowed to pass, since the deformation parameter depends on time.
For the Type 1 $q$-Gaussian distribution, we obtained the linear Fokker-Plank equation from the diffusion equation and showed that in this case the diffusion coefficient depends on both position and time.
By obtaining the continuity equation, it was shown both that the flow depends on the $q$-deformed diffusion coefficient, as well as the flow's compliance with Fick's law.
We repeat the calculation for the case in which at $t=0$ the substance is released at a point other than zero.

We then applied the Type 2  $q$-Gaussian distribution to find solutions to the diffusion equation, and the deformed diffusion-decay equation in a finite region was obtained. 
The position- and time-dependent decay coefficient was found, which goes to zero as the deformation parameter approaches zero.
The deformed diffusion-decay equation thus obtained   is different from the ordinary diffusion-decay equation because the ordinary derivative with respect to $x$ is replaced by the deformed derivative ${\cal D}_x^q$. 
To search for the physical meaning of this deformed diffusion-decay equation, a discrete version of it
in which the discrete times have a uniform interval, while the discrete positions have a non-uniform interval is addressed.

Because the Type 2 $q$-Gaussian distribution is related to the $q$-translation invariant derivative ${\cal D}_x^q$, we can construct the  $q$-translation invariant  quantum mechanics with a help of  ${\cal D}_x^q$. In such a theory, Type 2 $q$-Gaussian distribution can replace the ordinary Gaussian distribution in formulating the wave packet theory corresponding to the $q$-translation invariant  quantum mechanics with spatial derivative ${\cal D}_x^q$. Besides, Type 2 $q$-Gaussian distribution can be used in the wave function for the ground state in the $q$-translation invariant  quantum mechanics with spatial derivative ${\cal D}_x^q$.

We believe that the results found in this work are interesting and open the door to the use of this type of $q$-Gaussian functions for the analysis of some of the multiple physical phenomena that have been described in the introductory section and about which we are already working.

\section*{Acknowledgements}
The authors would like to thank the anonymous  referee for his/her valuable comments and suggestions.
This research was supported by the Q-CAYLE project, funded by the European Union-Next Generation UE/MICIU/Plan de Recuperacion, Transformacion y Resiliencia/Junta de Castilla y Leon (PRTRC17.11), and also by RED2022-134301-T and PID2023-148409NB-I00, financed by MI-CIU/AEI/10.13039/501100011033.  
Financial support of the Department of Education of the Junta de Castilla y León and FEDER Funds is also gratefully acknowledged (Reference: CLU-2023-1-05).


\begin{thebibliography}{99}
\expandafter\ifx\csname url\endcsname\relax
  \def\url#1{\texttt{#1}}\fi
\expandafter\ifx\csname urlprefix\endcsname\relax\def\urlprefix{URL }\fi
\expandafter\ifx\csname href\endcsname\relax
  \def\href#1#2{#2} \def\path#1{#1}\fi
	
\bibitem{TsallisJSP1988}  C. Tsallis, J. Stat. Phys. {\bf 52}  (1988) 479.


\bibitem{TsallisPA2004}
	C. Tsallis, Phys. A: Stat. Mech. Appl. {\bf  340} (2004) 1.
	
\bibitem{TsallisPTPS2006}
	C. Tsallis, Prog. Theor. Phys. Suppl. {\bf  162} (2006) 1.
	
\bibitem{Gell-Mann2004}
M. Gell-Mann, and C. Tsallis (Eds.),  Nonextensive Entropy -- Interdisciplinary Applications, Oxford University Press, New York, 2004.	

\bibitem{Frisch1995}
U. Frisch, Turbulence: The Legacy of A. N. Kolmogorov, Cambridge University Press, Cambridge, 1995.

\bibitem{BurlagaApJL2020}
L. F. Burlaga, N. F. Ness, D. B. Berdichevsky, L. K. Jian, J. Park, and A. Szabo, Astrophys. J. Lett. {\bf  901} (2020) L2.

\bibitem{WitkovskyMetrol2023}
V. Witkovsk\'y, Metrology {\bf  3} (2023) 222.

\bibitem{SicuroAP2015}
G. Sicuro, P. Tempesta, A. Rodr\'{i}guez, and C. Tsallis, Ann. Phys. (N. Y.) {\bf  363} (2015) 316.

\bibitem{daSilvaEPL2023}
C. V. da Silva, M. M. F. Nepomuceno, and D. B. de Freitas, Europhys. Lett., {\bf 141} (2023) 59002.

\bibitem{SekaniaPA2018}
M. Sekania, W. H. Appelt, D. Benea, H. Ebert, D. Vollhardt, and L. Chioncel, Phys. A: Stat. Mech. Appl. {\bf  489} (2018) 18.

\bibitem{DinhEPJB2018}
P. M. Dinh, L. Lacombe, P.-G. Reinhard, \'{E}. Suraud, and M. Vincendon, Eur. Phys. J. B {\bf  91} (2018) 246.  

\bibitem{Matsuzoe2021}
H. Matsuzoe, and A. Takatsu, Gauge freedom of entropies on $q$-Gaussian measures,  Progress in Information Geometry: Theory and Applications. Cham: Springer International Publishing, 2021, p. 127.

\bibitem{TirnakliPD2022}
U. Tirnakli, M. Mauricio, and C. Tsallis, Phys. D: Nonlinear Phenom. {\bf  431} (2022) 133132.

\bibitem{PajaresEPJA2023}
C. Pajares, and J. E. Ram\'{i}rez, Eur. Phys. J. A {\bf  59} (2023) 250.
\bibitem{KotaAP2022}
V. K. B. Kota, and M. Vyas, Ann. Phys. {\bf  446} (2022) 169131.
\bibitem{LiApJL2023}
X.-J. Li,  and Y.-P. Yang, Astrophys. J. Lett. {\bf  955} (2023) L34.



\bibitem{CuradoJPA1991} E. Curado, and C. Tsallis, J. Phys.  A: Math. Gen.  {\bf 24} (1991) L69.

\bibitem{Naudts2011} J. Naudts, Generalized Thermostatistics, Springer, 2011.

\bibitem{NobrePRL2022} F. Nobre, M. Rego-Monteiro, and  C. Tsallis, Phys. Rev. Lett.  {\bf 106} (2011) 140601.

\bibitem{DinizBJMBR2010}   P. Diniz, L. Murta-Junior, D. Brum, D. de Araujo, and A. Santos, Braz. J. Med. Biol. Res.  {\bf 43} (2010) 77.

\bibitem{FerriPA2010} G. Ferri, M.Savio, and A. Plastino, Physica A  {\bf 389} (2010) 1829.

\bibitem{EsquivelApJ2010} A. Esquivel, and  A. Lazarian, Astrophys. J.  {\bf 710} (2010) 125.

\bibitem{SotolongoGrauPRL2010} O. Sotolongo-Grau, D. Rodriguez-Perez, J. Antoranz, and O. Sotolongo-Costa, Phys. Rev. Lett.  {\bf 105} (2010) 158105.

\bibitem{AfsarPRE2010} O. Afsar, and  U. Tirnakli, Phys. Rev. E  {\bf 82} (2010) 046210.

\bibitem{ConroyPRD2008} J. Conroy, and  H. Miller, Phys. Rev. D  {\bf 78} (2008) 054010.

\bibitem{PlastinoPLA1993} A. R. Plastino, and A. Plastino, Phys. Lett. A  {\bf 174} (1993) 384.

\bibitem{RigoPLA2000} A. Rigo, A.R. Plastino, M. Casas, and  A. Plastino, Phys. Lett. A  {\bf 276} (2000) 97.

\bibitem{AssisJMP2005} P. Assis, L. da Silva, E. Lenzi, L. Malacarne, and  R. Mendes, J. Math. Phys.  {\bf 46} (2005) 123303.

\bibitem{PlastinoAP1997} A. R. Plastino, and  C. Anteneodo, Ann. Phys.  {\bf 255} (1997) 250.

\bibitem{ManzanoNJP2010} D. Manzano, R. Yanez, and  J.S. Dehesa, New J. Phys.  {\bf 12} (2010) 0.

\bibitem{NivanenRMP2003} L. Nivanen, A. Le Mehaute, and Q. Wang, Rep. Math. Phys.  {\bf 52} (2003) 437.

\bibitem{BorgesPA2004} E. Borges, Physica  A  {\bf 340} (2004) 95.

\bibitem{JacksonMM1909} F. Jackson, Mess. Math.  {\bf 38} (1909) 57.

\bibitem{ArikJMP1976} M. Arik, and D.Coon, J. Math. Phys.  {\bf 17} (1976) 524.

\bibitem{MacfarlaneJPAMG1989} A. Macfarlane, J. Phys. A: Math. Gen.  {\bf 22} (1989)  4581.

\bibitem{BiedenharnJPAMG1989} L. Biedenharn, J. Phys. A: Math. Gen.  {\bf 22} (1989) L873.

\bibitem{ChungFdP2019}  W. Chung, and H. Hassanabadi, Fortschr. Phys.  {\bf 67} (2019) 1800111.

\bibitem{ChungIJMP2019} W. Chung, and H. Hassanabadi,  Int. J.  Mod. Phys. A {\bf 2}  (2019) 1950177.

\bibitem{NaudtsJIPAM2004} J. Naudts, J. Ineq. Pure Appl. Math.  {\bf 5} (2004) 102.

\bibitem{NaudtsEpL2005} J. Naudts, Europhys. Lett.  {\bf 69} (2005) 719.

\bibitem{NaudtsOSID2005}  J. Naudts, Open Syst. Inf. Dyn.  {\bf 12} (2005) 13.

\bibitem{NaudtsPA2006} J. Naudts, Physica A  {\bf 365} (2006) 42.

\bibitem{NaudtsE2008} J. Naudts  Entropy  {\bf 10} (2008) 131.

\bibitem{NaudtsCEJP2009} J. Naudts, Cent. Eur. J. Phys.  {\bf 7} (2009) 405.

\bibitem{Grunwald2004} P. D. Gr\"{u}nwald, and A. P. Dawid,  Ann. Stat.  {\bf 32} (2004) 1367.

\bibitem{Eguchi2006} S. Eguchi, Sugaku Expositions  {\bf 19} (2006) 197.


\bibitem{AbeJPA2003}   S. Abe, J. Phys. A: Math. Gen.  {\bf 36} (2003) 8733.

\bibitem{HanelPA2007} R. Hanel, and S. Thurner, Physica A  {\bf 380} (2007) 109.

\bibitem{Topsoe2007} F. Tops{\o}e,  AIP Conference Proceedings {\bf  965} (2007) 104.

\bibitem{Topsoe2009} F. Tops{\o}e, J. Global Optim.  {\bf 43} (2009) 553.


\bibitem{UmarovMJM2008} S. Umarov, C. Tsallis, and S. Steinberg, Milan J. Math.  {\bf 76} (2008) 307.

\bibitem{dOnofrio2013} A. d'Onofrio (ed.),  Bounded Noises in Physics, Biology, and Engineering, Birkhauser, 2013.


\bibitem{Fick1855}  A. Fick, Ann. Phys. (Berlin) {\bf  94} (1855) 59.


\bibitem{PapNSJM1993} E. Pap,  Novi Sad J. Math.  {\bf 23}  (1993) 145.

\bibitem{ChungRMP2020} W. S. Chung, and H. Hassanabadi, Rep. Math. Phys.  {\bf 85} (2020) 305.

\bibitem{new} C. Tsallis,   Introduction to nonextensive statistical mechanics: approaching a complex world, Springer, New York, 2023, pp. 1-2.

\end{thebibliography}
\end{document}